\documentclass[twocolumn,prl]{revtex4}

\usepackage{epsf}

\begin{document}
\preprint{}
%


\title{Classical simulation of quantum algorithms using the tensor product representation}
\author{A. Kawaguchi$^{1,3}$, K. Shimizu$^1$, 
Y. Tokura$^{1,3}$ and N. Imoto$^{1,2,3,4}$}
\address{
$^1$
NTT Basic Research Laboratories, NTT Corporation, 
3-1 Morinosato-Wakamiya, Atsugi, Kanagawa 243-0198 
\\
$^2$
Department of Materials Engineering Science, 
Osaka University, Toyonaka, Osaka 560-8531, Japan
\\
$^3$
SORST Research Team for Career Electronics, JST
\\
$^4$
CREST Research Team for Photonic Information Processing, JST
}
\date{\today}
\begin{abstract}

Using the tensor product representation in the density matrix renormalization group, we show that a quantum circuit of Grover's algorithm, which has one-qubit unitary gates, generalized Toffoli gates, and projective measurements, can be efficiently simulated by a classical computer. 
It is possible to simulate quantum circuits with several ten qubits.

\end{abstract}

\pacs{PACS numbers: }

\maketitle



Quantum computation is currently attracting much interest \cite{Shor,Grover,Nielsen}. 
Shor's factorization algorithm \cite{Shor}, which finds the prime factorization of an $n$-bit integer in the polynomial time of $n$, proved the power of quantum computation . 
Grover's search algorithm \cite{Grover}, which has complexity $O(\sqrt N)$ instead of $O( N)$ classical steps, where $N$ is the database size, 
provides proof that quantum computers are faster than classical ones for unstructured database searches. 
Moreover, it is expected to have applications to other important NP problems.

Actual physical systems are affected by various decoherence and operational errors, which make it difficult to analyze the properties of quantum circuits. 
Therefore, it is very important to simulate a quantum computer realized with physical systems by a classical computer.  
However, simulating large quantum circuits with a classical computer requires enormous computational power because the number of quantum states increases exponentially as a function of the number of qubits. 
Recently, Vidal presented an efficient method for simulating quantum computations with one-qubit unitary gates and neighboring two-qubit unitary gates \cite{Vidal1}. 
Before that, White had introduced a numerical technique to study the properties of relatively large-scale one-dimensional (1D) quantum systems, 
which is called the density matrix renormalization group (DMRG) \cite{White}. 
Vidal's idea and DMRG strongly overlap.  
As an interesting combination of the above two methods, a time-dependent DMRG method has been presented \cite{Vidal2, White2}.



In this paper, we present a new simulation method that uses the graphical representation of the tensor product state \cite{Nishino}. 
With this method, we can take advantage of arbitrary one- and two-qubit operators, which include projection operators, and can execute 
the projective measurement in an efficient way. 
In addition, generalized Toffoli gates (C$^k$-NOT) can be used without decomposition in terms of elementary gates. 
As an example of the simulation, we will show that Grover's algorithm can be efficiently executed by a classical computer. 
Here, to implement Grover's algorithm concretely, we assume that the Oracle consists of generalized Toffoli gates. 
Using this simulation method, it becomes possible to simulate various quantum circuits with several ten qubits.


According to the tensor product decomposition, any wave function of pure state $\Psi_0$ with $L$ qubits can be decomposed to $L$ tensors.
\begin{eqnarray}
&&\hspace{-1.3cm}
\Psi_0(i_1,i_2,\cdots, i_L)=
\sum_{x_2=1}^{m_2} \cdots \sum_{x_{L}=1}^{m_{L}}
\nonumber \\
&&\hspace{-1.0cm}
U_1(i_1,x_1,x_2)U_2(i_2,x_2,x_3)\cdots U_{L}(i_L,x_L,x_{L+1}), 
\label{EQ_TP1}
\end{eqnarray}
where $i_{\alpha}$ represents the $\alpha$-th qubit.
The right side consists of the product of $L$ tensors, and $x_ {\alpha}(=1, \cdots m_ {\alpha})$ are new variables related to the entanglement between neighboring qubits \cite{Vidal1}. 
Here, $x_1=x_{L+1}=1$ ($m_1=m_{L+1}=1$) for the pure state. 
This representation contains $O (m^2L) $ components, where $m$ is the maximum value of $m_ {\alpha}$.

\begin{figure}[tb]
\begin{center}
\vspace{-1.5cm}
\leavevmode \epsfxsize=100mm 
\epsffile{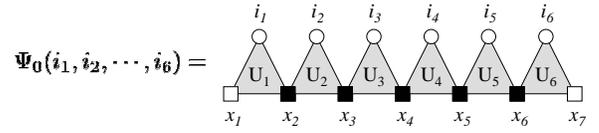}
\vspace{-11.0cm}
\end{center}
\caption{ Graphical representation of the tensor product state for $L=6$ in Eq. (\ref{EQ_TP1}). 
}
\label{TP1}
\end{figure}

In Fig. \ref{TP1}, we show the graphic representation of Eq. (\ref{EQ_TP1}) with $L=6$  \cite {Nishino}.
In this graph, we represent tensors $U_{\alpha}$ as triangles, and each vertex corresponds to $i_{\alpha}$ (circles) and $x_{\alpha}$ (squares).
We use  black marks in the same manner as Einstein's summation convention. 
The representation implies that a summation is taken over the corresponding index when an index is repeated once in the same term. 
Here, all $m_{\alpha}$ become $m_{\alpha}=1$ for the product states. 
Therefore, tensors $U_{\alpha}(i_{\alpha},x_{\alpha},x_{\alpha+1})$ are reduced to vectors $U_{\alpha}(i_{\alpha})$.
For example, in the case of a superposition state, $1/2^ {L/2} \sum_ {i=1} ^ {2^L} |i\rangle$, then $U_{\alpha} = (1/\sqrt {2}, 1/\sqrt {2}) ^ {T} $. 
Also, when a state is $ | 00\cdots0\rangle$, then $U_{\alpha}=(1,0)^{T}$.
In this paper, we use the $| 00\cdots0\rangle$ state as the initial state to simulate quantum circuits.


\begin{figure}[tb]
\begin{center}
\vspace{-1.5cm}
\leavevmode \epsfxsize=100mm 
\epsffile{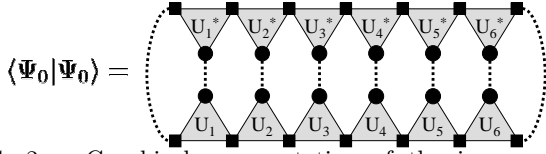}
\vspace{-11.0cm}
\end{center}
\caption{
Graphical representation of the inner product $\langle \Psi_0 | \Psi_0 \rangle$.
}
\label{TP3}
\end{figure}

As a simple example, we explain a calculation process for the inner product. 
Fig. \ref{TP3} is the graphical representation of the inner product $\langle \Psi_0 | \Psi_ 0 \rangle$ for $L=6$, 
where dotted lines imply that a summation is taken over the connected variable.
We calculate $\langle \Psi_0 | \Psi_ 0 \rangle$ as follows: 
\begin{eqnarray}
&&X_6(x_6,x_6^{\prime})=\sum_{i_6,x_7}
          U_6(i_6,x_6,x_7)U_6^*(i_6,x_6^{\prime},x_7)
\\
&&Y_5(i_5,x_5,x_6^{\prime})=\sum_{x_6}
         X_6(x_6,x_6^{\prime}) U_5(i_5,x_5,x_6)
\\
&&X_5(x_5,x_5^{\prime})=\sum_{i_5,x_6^{\prime}}
        Y_5(i_5,x_5,x_6^{\prime}) U_5^*(i_5,x_5^{\prime},x_6^{\prime})
\\
&&\hspace{1.0cm} \vdots \hspace{2.5cm} \vdots \nonumber \\
&&X_1(x_{1},x_{1}^{\prime})=\sum_{i_1,x_2^{\prime}}
    Y_1(i_1,x_1,x_2^{\prime})  U_1^*(i_1,x_1^{\prime},x_2^{\prime}). 
\label{EQ_Inner}
\end{eqnarray}
Here, $X_1(x_{1},x_{1}^{\prime})$ has the scalar quantity because $x_{1}=x_{1}^{\prime}=1$; therefore, $\langle \Psi_0 | \Psi_ 0 \rangle= X_1(1,1)$. 
The calculation of $\langle \Psi_0 | \Psi_ 0 \rangle$ for $L$ qubits requires $O (Lm^3)$ multiplications.



\begin{figure}[tb]
\begin{center}
\vspace{-0.5cm}
\leavevmode \epsfxsize=90mm 
\epsffile{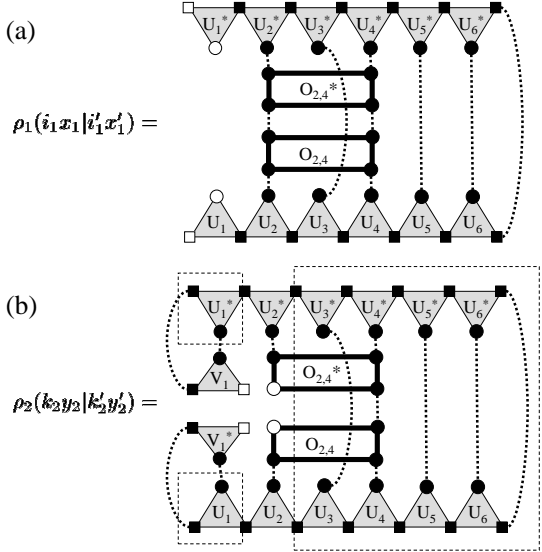}
\vspace{-5.0cm}
\end{center}
\caption{Graphical representation of 
$\rho_1(i_1,x_1|i_1^{\prime},x_1^{\prime})$ and 
$\rho_2(k_2,y_2|k_2^{\prime},y_2^{\prime})$. 
}
\label{TP2}
\end{figure}

To simulate a quantum circuit one step further, we have to obtain the tensor product decomposition of the next state 
$|\Psi_1\rangle=O_p |\Psi_0\rangle$,  
\begin{eqnarray}
&&\hspace{-1.3cm}
\Psi_1(i_1,i_2,\cdots, i_L)=
\sum_{y_2=1}^{m^{\prime}_2}
 \cdots \sum_{y_{L}=1}^{m^{\prime}_{L}}
\nonumber \\
&&\hspace{-1.0cm} 
V_1(i_1,y_1,y_2)V_2(i_2,y_2,y_3)\cdots V_{L}(i_L,y_L,y_{L+1}), 
\label{EQ_Psi1}
\end{eqnarray}
where $y_1=y_{L+1}=1$. 
Here, we assume $L=6$ and the operator is $O_p= O_{2,4}(k_2,k_4| i_2,i_4)$. 
First, we consider the reduced density matrix $\rho_1(i_1,x_1| i_1^{\prime},x_1^{\prime})$ in Fig. \ref{TP2}(a) to get $V_1$ and $m^{\prime}_2$.
The order of the calculation becomes slightly complicated, but the calculation can be performed similarly to that for the inner product. 
Then, we diagonalize the $2\times2$ density matrix $\rho_1(i_1,x_1| i_1^{\prime},x_1^{\prime})$, 
\begin{eqnarray}
&&\sum_{i_1,x_1,i_1^{\prime},x_1^{\prime}}
V_{1}^{*}(i_1,x_1,y_2)
\rho_1(i_1,x_1|i_1^{\prime},x_1^{\prime})
V_{1}(i_1^{\prime},x_1^{\prime}, y_2^{\prime})
\nonumber \\
&&\hspace{1.5cm} =\delta_{y_2,y_2^{\prime}} \lambda_{y_2}.
\label{Diag}
\end{eqnarray}
Here, all eigenvalues $\lambda_{y_2}$ have positive values in the decreasing order ( $\lambda_1 \ge \lambda_2 $ ). 
Then, we can rewrite $V_1(i_1,x_1, y_2)=V_1(i_1,y_1, y_2)$ since $x_1=y_1=1$ ($m_1=m^{\prime}_1=1$). 
Furthermore, we choose $m^{\prime}_2$ as $m^{\prime}_2=2$ when $\lambda_2\ne0$, and $m^{\prime}_2=1$ when $\lambda_2=0$. 
When DMRG is used in the analysis of 1D quantum lattice systems, such as the Heisenberg model, it is preferable to cut $m^{\prime}_{\alpha}$ that the $(m_{\alpha}+1)$-th eigenvalue of a density matrix becomes small enough. 
This corresponds to an irreversible compression that disposes of the unimportant information, and such compression is very effective.
However, because quantum calculations, such as Grover's and Shor's algorithms, tend to treat discrete wave functions, important information is left out if we make $m_{\alpha}$ too small. 
Therefore, we determine $m^{\prime}_{\alpha}$ such that the eigenvalue of a density matrix becomes 0 in this paper. 
Namely, $m^{\prime}_{\alpha}$ corresponds to the Schmidt rank of the density matrix $\rho_{\alpha-1}$ and the compression corresponds to the reversible compression that the original information is fully maintained.
Next, we use $V_1$ obtained previously and calculate the density matrix $\rho_2(k_2,y_2| k_2^{\prime},y_2^{\prime})$ as Fig. \ref{TP2}(b).
Here, the qubit variable changes from $i_2$ to $k_2$  because of $O_{2,4}$. 
Then, $V_2$ and $m^{\prime}_3$ can be obtained by diagonalizing $\rho_2$ 
in the same manner as obtaining $V_1$ and $m^{\prime}_2$. 
By repeating the above-mentioned operations, we can get all $V_{\alpha}$ and $m^{\prime}_{\alpha}$.

Next, we evaluate the computational complexity 
for one-qubit and two-qubit operations and show 
that the complexity is $O(Lm^3)$ for a projection operator.
The calculation for obtaining $\rho_{1}$ requires $O(Lm^3)$ multiplications, and  $O(m^3)$ multiplications are necessary in order to diagonalize it. 
In the calculation for preparing $\rho_{\alpha}$ ($\alpha >1$), some parts (ex. dashed boxes in Fig. \ref{TP2}) are already obtained through the previous calculation. This reduces the calculation for preparing $\rho_{\alpha}$ to $O(m^3)$ multiplications. 
Therefore, the total number of multiplications is $O(Lm^3)$ for getting all $V_{\alpha}$ and $m^{\prime}_{\alpha}$. 
Note that, for a one-qubit unitary $O_{j}$ and a neighboring two-qubit unitary operator $O_{j,j+1}$, the calculation complexities can be reduced to $O(m^2)$ and $O(m^3)$, respectively \cite{Vidal1}. 
In the case of the one-qubit unitary operator $O_{j}$, 
$V_j(k_j,x_j,x_{j+1} )=\sum_{i_j}O_{j}(k_j,i_j) U_{i}(i_j,x_j,x_{j+1} )$ 
  and $V_{\alpha}=U_{\alpha}$ for $\alpha\ne j$. 
Also, for the unitary operator $O_{j,j+1}$, the computational complexity for obtaining the next state becomes $O(m^3)$ by using $V_{\alpha}=U_{\alpha}$ ( $\alpha\ne j,j+1$ ). 
However, for the projection operator, all $U_{\alpha}$ may change due to the collapse of the wave function. 
Thus, the computational complexity is $O(Lm^3)$ for a projection operator. 
We will explain the method of projective measurement later.


Furthermore, we can use the C$^k$-NOT gate with $k$ controlled qubits and one target qubit without decomposing in terms of elementary gates. 
Though the C$^k$-NOT gate can be decomposed into single qubit unitary gates and C-NOT gates \cite{Barenco}, the C$^k$-NOT gate is very useful for constructing a quantum circuit. 
As a simple example, we show the case of a C$^2$-NOT operation for $L=4$, which has  two controlled qubits ($i_1$ and $i_2$) and  one target qubit ($i_4$).
In this case, all elements of $\Psi_0(i_1,i_2,i_3, 1)$ are replaced with $\Psi_0(i_1,i_2,i_3, 0)$ when both $i_1$ and $i_2$ are  $|1\rangle$. 
Therefore, we can represent the next state $\Psi_1(i_1,i_2,i_3, i_4)$ as follows.
\begin{eqnarray}
&&\hspace{-1.3cm}
\Psi_1(i_1,i_2,i_3, i_4)=\Psi_0(i_1,i_2,i_3, i_4)
\nonumber \\
&&-\Psi_0(1,1,i_3, i_4)
+\Psi_0(1,1,i_3, \bar{i_4}). 
\label{EQ_TP4}
\end{eqnarray}
Here, $\bar{i_4}$ means the reverse of $i_4$.  
The $i_1$ and $i_2$ in the second and third term are fixed to $1$. 
We can make the reduced density matrixes using the tensor product representation and obtain $V_{\alpha}$ and $m^{\prime}_{\alpha}$ one after another.
In this way, we can use C$^k$-NOT with $O(L m^3)$ multiplications. 
Namely, we can use almost all operations appearing in quantum circuits with less than $O(L m^3)$ multiplications.


\begin{figure}[tb]
\begin{center}
\vspace{-0.5cm}
\leavevmode \epsfxsize=90mm 
\epsffile{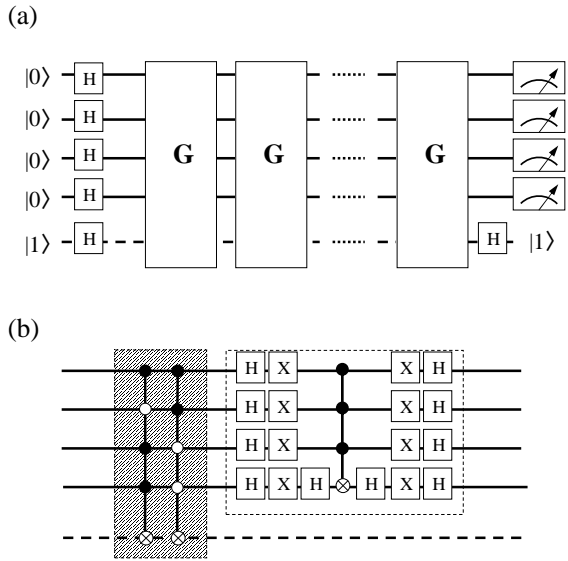}
\vspace{-4.5cm}
\end{center}
\caption{ 
Full Grover's algorithm (a) and Grover operator (b) 
for $n=4$ with the number of solutions $t=2$.   
In this example, correct solutions are $|1011 \rangle$ and $|1100 \rangle$.
}
\label{Grover1}
\end{figure}

\begin{figure}[tb]
\begin{center}
\vspace{-0.0cm}
\leavevmode \epsfxsize=90mm 
\epsffile{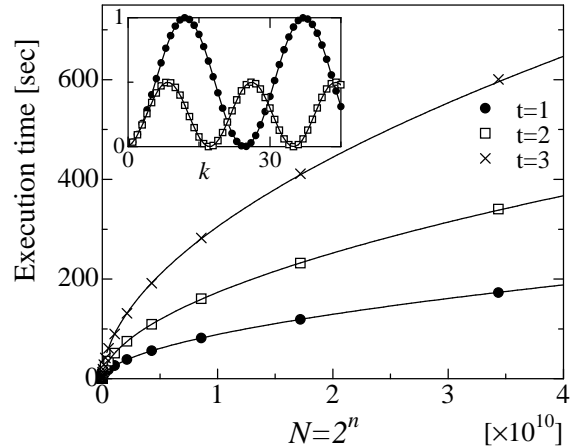}
\vspace{-7.0cm}
\end{center}
\caption{Execution time of a classical computer for Grover's algorithm up to $n=35$.  
The inset shows the amplitude of a correct solution for $n=8$.
Filled circles and open squares indicate $t=1$ and $t=2$, respectively.
The execution time indicates the computational time 
from $k=0$ to the first maximum, $k\sim \pi\sqrt{N/t}/4$, in the inset.
}
\label{Time}
\end{figure}

 Grover's search algorithm involves repeated applications of a quantum subroutine, which is called the Grover operator $G$. 
In Fig. \ref{Grover1}, we illustrate a configuration example of $n=4$ with the number of solutions $t=2$.  
Here, $H$ and $X$ represent the Hadamard gate,  
${ \displaystyle
H=\frac{1}{\sqrt 2}\left(
    \begin{array}{cc}
      1 & 1 \\
      1 & -1
    \end{array}
  \right)  }$, 
and the NOT gate,  
$ { \displaystyle
X=\left(
    \begin{array}{cc}
      0 & 1 \\
      1 & 0
    \end{array}
  \right)   }$. 
In this study, we assume that the Oracle [shaded part in Fig. \ref{Grover1}(b)] consists of $t$ C$^n$-NOT gates. 

When we apply $k$ times of $G$ to a superposition state, the state is represented as 
\begin{eqnarray}
&& G^k|\Psi\rangle = \cos \left(\frac{2k+1}{2}\theta \right)|\alpha\rangle 
                + \sin \left(\frac{2k+1}{2}\theta \right)|\beta\rangle , 
\label{Goperator1}
\end{eqnarray}
where $\theta=2\arcsin\sqrt{t/N} $ and 
\begin{eqnarray}
|\alpha\rangle &=& \frac{1}{\sqrt{N-t}} {\sum_x}'' |x \rangle \\
|\beta\rangle &=& \frac{1}{\sqrt{t}} {\sum_x}'  |x \rangle. 
\label{Goperator2}
\end{eqnarray}
Here, ${\sum_x}'$ indicates a sum over all $x$ that are solutions, and ${\sum_x}''$ indicates a sum over all $x$ that are not solutions. 
In the inset of Fig. \ref{Time}, we show the amplitudes of a correct state for $t=1,2$ and $n=8$. 
For convenience, we have fixed one of the solutions in $|11\cdots1\rangle$ and have plotted the amplitude of $|11\cdots1\rangle$. 
Solid lines indicate analytical lines, $|\sin \left(\frac{2k+1}{2}\theta \right)|^2/t$.  
These results confirm that tensor product representation can simulate Grover's algorithm precisely.

A correct solution is provided  by measuring qubits when $k\sim \pi\sqrt{N/t}/4$.
In our numerical simulation, a projective measurement is performed as follows.
First we calculate an expectation value $\langle\Psi |\hat{n}_i|\Psi\rangle$ of $i$-th qubit and generate a random number $r$ ($0\leq r \leq 1$) to consider the probabilistic nature of quantum measurement, 
where ${ \displaystyle
\hat{n}_i=\left(
    \begin{array}{cc}
      0 & 0 \\
      0 & 1
    \end{array}  \right)  }$. 
We apply to the $i$-th qubit 
\begin{equation}
\left\{     
\begin{array}{cc}
P_0=| 0 \rangle \langle 0 |
&  \hspace{0.5cm} {\rm if }  \langle\Psi |\hat{n}_i|\Psi\rangle \leq r  \\
P_1=| 1 \rangle \langle 1 |
&  \hspace{0.5cm} {\rm if }  \langle\Psi |\hat{n}_i|\Psi\rangle > r .
\end{array}
\right.
\label{Eq_Proj}
\end{equation}


In Fig. \ref{Time}, we show the execution time of the quantum circuit illustrated in Fig \ref{Grover1} up to $n=35$ using an Intel Xeon 2.4GHz processor. 
What has to be noticed is that the computation time increases when the number of answers increases. 
This behavior is completely opposite to that for an ordinary search problem. 
Here, we discuss the computational complexity of a classical computer. 
The circuit possesses $O(\sqrt{N/t}\log_2 N)$ one-qubit unitary gates, 
$O(t \sqrt{N/t})$ C$^k$-NOT gates, and $\log_2 N$ projective measurements. 
The calculation complexities per one gate are $O(m^2)$, $O(m^3 \log_2 N)$, and $O(m^3 \log_2 N)$, respectively. 
Because the total computational complexity is mainly dependent on C$^k$-NOT gates, it becomes $O(\sqrt{tN} m^3 \log_2 N)$.  
We performed the fitting of $A\sqrt{N}(\log N)$ with a fitting parameter $A$ (solid lines in Fig. \ref{Time}), 
and these fitting lines are consistent with the simulation results.  
Therefore, we see that $m$ is not dependent on the qubit number $n(=\log_2 N)$. 
Actually, we have found that $m$ has the relationship $m\leq t+1$. 
The total computational complexity for finding one correct state can be expressed by $O(\sqrt{N} t^{3.5} \log_2 N)$. 
Incidentally, the required memory capacity  is $O(t^{2} \log_2 N)$.
If $t$ is much smaller than $N$, this computational complexity is approximately equal to that for a quantum computer. 
In this way, we have found that the tensor product representation is very useful in simulating  Grover's algorithm.


%
%

In summary, we have demonstrated the classical simulation of a quantum circuit by using the graphical representation of the tensor product state. 
This method can take advantage of almost all gate operations,  which include projection operators and the generalized Toffoli gates, with less than $O(Lm^3)$ multiplications. 
As a result of the simulation of Grover's algorithm, we have found that $m$, which determines the  computational time and the memory size in a classical computer, is related to the number of solutions.
This simulation method enables us to simulate various quantum circuits with several ten qubits. 
Therefore, we expect that this method will be helpful in analyzing quantum circuits with decoherence and operational errors.
In particular, since this simulation method is quite effective for a one-dimensional quantum system, it will be useful for analyzing several important experimental approaches,
which include a one-dimensional chain of quantum dots \cite{Dot} 
and the all-silicon quantum computer \cite{Kane, Silicon}.

We are grateful to  F. Morikoshi, N. Kunihiro and Y. Tokunaga  
for valuable discussions about the quantum computation. 
We also thank K. Okunishi and T. Nishino for discussions about the graphical representation of tensor product state.


\begin{thebibliography}{99}


\bibitem{Nielsen} 
M. A. Nielsen and I. L. Chuang,{\it Quantum Computation and Quantum Information} (Cambridge University Press, Cambridge, 2000)

\bibitem{Grover} 
L. K. Grover, PRL {\bf 79}, 325 (1997).

\bibitem{Shor}
P. W. Shor, in Proceeding of 35th Annual Symposium on Foundation of 
Computer Science, IEEE Computer Society Press, Los Alamitos, CA, 124 (1994).

\bibitem{Barenco}
A. Barenco, C.H. Bennett, R. Cleve, D.P. DiVincenzo, 
N. Margolus, P. Shor, T. Sleator, J. A. Smolin and H. Weinfurter,
  Phys. Rev. A {\bf 52}, 3457 (1995).


\bibitem{White}
S. R. White, 
Phys. Rev. Lett. {\bf 69}, 2863 (1992); 
Phys. Rev. B {\bf 48}, 10345 (1993). 



\bibitem{Nishino}
See for instance T. Nishino, T. Hikihara, K. Okunishi and Y. Hieida, 
International Journal of Modern Physics B {\bf 13} 1 (1999).


\bibitem{Vidal1}
G. Vidal, 
Phys. Rev. Lett. {\bf 91}, 147902 (2003). 
\bibitem{Vidal2}
G. Vidal, 
Phys. Rev. Lett. {\bf 93}, 040502 (2004); 
 A. J. Daley, C. Kollath, U. Schollwoeck, G. Vidal, cond-mat/0403313. 
\bibitem{White2}
S. R. White and A. E. Feiguin, 
Phys. Rev. Lett. {\bf 93}, 076401 (2004). 






\bibitem {Dot}
D. Loss and D. P. DiVincenzo, 
Phys. Rev. A {\bf 57}, 120 (1998).

\bibitem {Kane}
B. E. Kane, 
Nature {\bf 393}, 133 (1998).

\bibitem {Silicon}
T. D. Ladd, J. R. Goldman, F. Yamaguchi, Y. Yamamoto, 
E. Abe, and K. M. Itoh, 
Phys. Rev. Lett. {\bf 89}, 17901 (2002).




\end{thebibliography}
\end{document}